\documentclass[conference]{IEEEtran}
\IEEEoverridecommandlockouts
\usepackage{cite}
\usepackage{amsmath,amssymb,amsfonts}
\usepackage{algorithmic}
\usepackage{graphicx}
\usepackage{textcomp}
\usepackage{xcolor}
\usepackage{amsmath}
\usepackage{comment}
\usepackage{float}
\usepackage{stfloats}
\restylefloat{table}
\usepackage{threeparttable}
\usepackage{tabularx}
\usepackage{booktabs}

\usepackage{cite}

\def\BibTeX{{\rm B\kern-.05em{\sc i\kern-.025em b}\kern-.08em
    T\kern-.1667em\lower.7ex\hbox{E}\kern-.125emX}}

\begin{document}

\makeatletter % changes the catcode of @ to 11
\newcommand{\linebreakand}{%
  \end{@IEEEauthorhalign}
  \hfill\mbox{}\par
  \mbox{}\hfill\begin{@IEEEauthorhalign}
}
\makeatother % changes the catcode of @ back to 12

\title{Towards Fair and Privacy Preserving Federated Learning for the Healthcare Domain}

\author{
    \IEEEauthorblockN{
    Navya Annapareddy \textsuperscript{\rm 2},
    Yingzheng Liu \textsuperscript{\rm 1},
    Judy Fox \textsuperscript{\rm 1,2}
    }

    \IEEEauthorblockA{
        \textsuperscript{\rm 1} Computer Science Department, University of Virginia \\
     \textsuperscript{\rm 2} School of Data Science, University of Virginia
      \\
      Charlottesville, USA \\
    Email : \{mi3se, yqx8es, yl4dt, wsw3fa, njd9e, cwk9mp\}@virginia.edu 
    }
}

\maketitle

\begin{abstract}
Federated learning (FL) enables data sharing in healthcare contexts where it might otherwise be difficult due to data-use-ordinances or security and communication constraints.  Distributed and shared data models allow models to become generalizable and learn from heterogeneous clients. While addressing data security, privacy, and vulnerability considerations, data itself is not shared across nodes in a given learning network. On the other hand, FL models often struggle with variable client data distributions and operate on an assumption of independent and identically distributed(IID) data. As the field has grown, the notion of fairness-aware federated learning (FAFL) mechanisms has also been introduced and is of distinct significance to the healthcare domain where many sensitive groups and protected classes exist. In this paper, we create a benchmark methodology for FAFL mechanisms under various heterogeneous conditions on datasets in the healthcare domain typically outside the scope of current federated learning benchmarks, such as medical imaging and waveform data formats. Our results indicate considerable variation in how various FAFL schemes respond to high levels of data heterogeneity. Additionally, doing so under privacy-preserving conditions can create significant increases in network communication cost and latency compared to the typical federated learning scheme.

\end{abstract}


\section{Introduction}
Privacy and security have been a growing concern with machine learning systems. Having huge amounts of data in the same place is a security risk. Therefore, many domains such as healthcare regulate the centralization of data. Addressing this concern of centrally located data and privacy, McMahan et al. contributed to the concept of federated learning (FL) and distributed architecture \cite{mcmahan2017communication}. In a federated learning system, decentralized clients load and train data on their own devices and communicate weights to a global server and is therefore ideal for the sensitive nature of clinical environments.

\begin{figure}[htbp]
\centerline{\includegraphics[width=8.6cm, height=6.5cm]{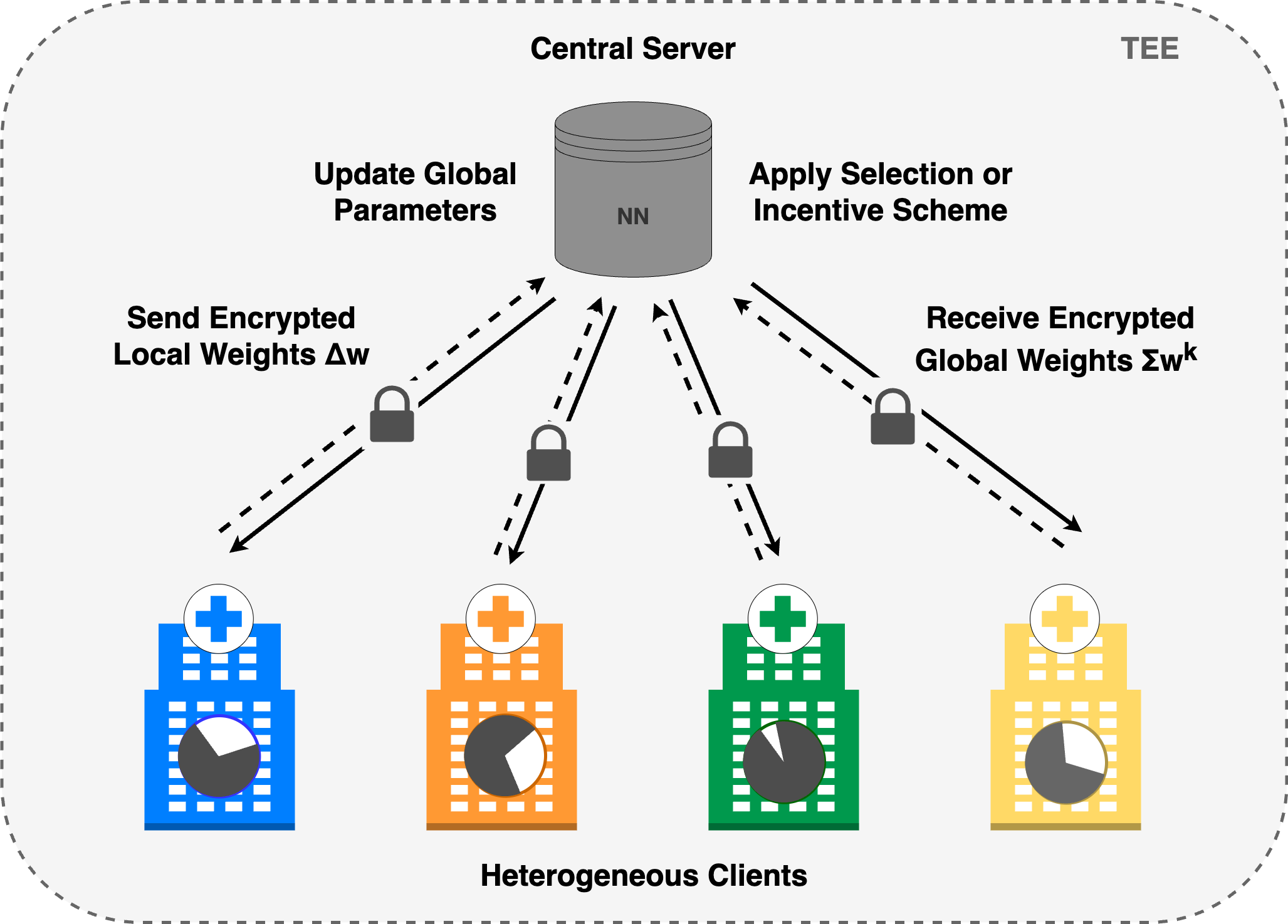}}
\caption{Discrete federated learning lifecycle with heterogeneous client data distributions and server level processes and interactions. After model training and inference, key client server interactions (CSIs) include encrypting and decryption local and global weights during model aggregation.}
\label{fig:fig1}
\end{figure}

Fairness within FL systems can refer to both individual and group parity, performance, and participation. FAFL can involve every stage of the FL lifecycle as seen in Fig. \ref{fig:fig1}, including selection and collaboration \cite{Shi}. Like with FL usage in healthcare, there is no standard usage of FAFL mechanisms and analysis is typically constrained to single methods. 

There exist many health data ordinances that ensure patient privacy and confidentiality, commonly through compliance fixtures \cite{ahmad2020fairness}.  Machine learning in healthcare typically involve multiple networks with differing communication and capacities  \cite{JAVAID202258}. Moreover, healthcare incorporates a wide variety of data, including those from minimally encrypted  instruments with heterogeneous capabilities or whose infrastructure is split between servers, leaving them vulnerable to hijacking or inversion \cite{sidechannel}. The lack of standard FL in healthcare usage remains a challenge as implementations remain context specific. The contributions of this paper will include jointly implementing FAFL mechanisms under an AES-GCM encryption scheme and proposing additional benchmarks on common healthcare datasets as described in Table \ref{table:kysymys}.

\section{Related Work} 
\subsection{Federated Learning Benchmarks}
Many traditional benchmarks of federated learning implementations on health datasets, such as signal datasets and clinical databases like MIMIC III, do not incorporate fairness mechanisms \cite{lee2020federated}. Several applications for differential privacy and encrypted computation have been separately proposed for health contexts, notably PySyft and Microsoft SEAL \cite{Pysyft_mimic}.  Unified benchmark systems like FedScale have an apparent lack of built in health datasets and encryption methods.
\subsection{Fairness Approaches} 
\subsubsection{Client Selection}
Client selection and filtering are stages of nodes selection based on on potential to improve the global model \cite{Shi}. Clients' resource constraints can determine its capacity to contribute to the global model, leading to clients being filtered out before the first round. Clients are rewarded or removed from the global model depending on  performance.

\subsubsection{Contribution Fairness}
Contribution evaluation typically consider fairness in node filtering and rewards. This approach is key for differential privacy, as individual features can be removed for deidentification. Participants are rewarded based on novel global contributions often by rewarding participants based on test accuracy of the client’s model \cite{Reputation}.

\begin{figure*}
\begin{center}
  \includegraphics[width=16.7cm, height=5.55cm]{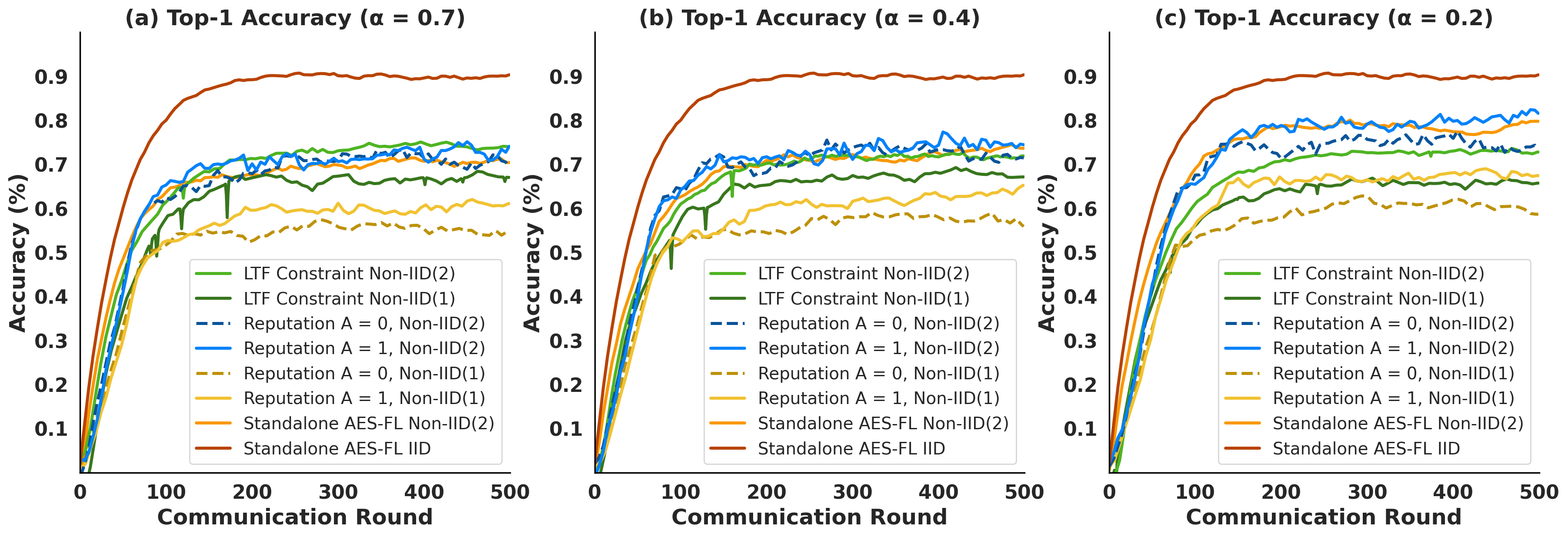}
  \caption{Benchmarks for inference on NIH Chest X-Ray Dataset with classification models and varying ratios of non-IID data in client data distribution as dictated by parameter alpha. Top-1 accuracy is reported for for standalone encrypted FL (AES-FL), long term client selection fairness constraint (LTF Constraint), reputation mechanism without adversarial agent (Reputation A = 0), and reputation mechanism with adversarial agent  (Reputation A = 1) schemes with different maximum unique labels per client. Non-IID(1) and non-IID(2) refer to enforcement of a 1-class and 2-class non-IID distribution respectively. Optimal scheme is long term fairness for alpha = 0.7 (\textbf{Left}) and reputation with adversarial agent for alpha = 0.4 (\textbf{Center}) and alpha = 0.2 (\textbf{Right}).
  }
  \label{fig:fig2}
\end{center}
\end{figure*}

\begin{table*}[!b]
\caption{Generalized formulation of fairness mechanisms. Client and global parameter set for the next time step are represented by $w^{t+1}_{k}$ and $w^{t+1}_{i}$. Unique terms include gradient regularization parameter $\gamma$ and group loss term $g$. We let $\Delta w^t_{i}$ be implicitly representative of a designated loss function $f_k(w^t_i)$ applied to weights at round $i$ for $K$ clients.}

\centering
\def\arraystretch{1.2}
\begin{tabular}{p{0.22\linewidth}p{0.26\linewidth}p{0.2\linewidth}p{0.24\linewidth}p{0.12\linewidth}}
\hline
\textbf{Fairness Approach} & \textbf{Implementation} & \textbf{Local Protocol} & \textbf{Global Protocol}\\
\hline
(1) Client Selection  & Long Term Fairness \cite{contribution} & $C^2 MAB(\eta, c^t_k, \Delta w^t_{i*})$  & $\frac{1}{K} {\Sigma}^K_{i=1} w^t_i$ if $ I^t_k = 1 $ \\
(2) Contribution Fairness  & Reputation Threshold \cite{Reputation}\cite{long}  & $ w^t_k + \Delta w^t_k + \Delta w^t_{i*}, r^t_k$ & $ {\Sigma}^R_{i=1} r^{t-1}_i w^t_i \times \gamma \big/ \lVert \Delta w^t_i \rVert $\\
(3) Group Fairness & Bounded Group Loss \cite{BoundedGroupLoss}  & $w^t_k -(\lambda^T r + \Delta w^t_{i*})$ & $ w^t_i + \frac{1}{K} {\Sigma}^K_{i=1} g_i^{t+1} $\\ 
(4) Agnostic Fairness & Agnostic Loss \cite{afl}  & $\min\max  
     L(\lambda^t_i, \Delta w^t_{i*}) $ &  $\frac{1}{K} {\Sigma}^K_{i=1} w^t_i, \frac{1}{K} {\Sigma}^K_{i=1} \lambda^t_i$\\
(5) Expectation Fairness & Incentive Sharing \cite{incentive}   & $\max V(s^t_k, \Delta w^t_{i*})^k$ & $\frac{1}{K} {\Sigma}^K_{i=1} w^t_i$, $\pi^t_i(w^t_i, s^t_i)$
    \\
\hline
\end{tabular}
\label{table:defs}
\end{table*}

\subsubsection{Group Fairness}
Group fairness refers in-group performance parity, such as basic assured accuracy. Huang et al. \cite{Huang} presented long-term client selection fairness constraints while Hu et al. satisfy global model fairness requirements through a conditional group loss, enforcing minimum fairness metrics but not requiring them to be equal \cite{BoundedGroupLoss}.  

\subsubsection{Agnostic Learning}
Agnostic loss based federated learning (AFL) models of fairness prevent overfitting by optimizing for variable client mixtures instead of assuming distribution parity between target and client. They have been noted struggle to generalize with large clients \cite{afl}.  

\subsubsection{Expectation and Regret Fairness}
Under a regret fairness scheme, nodes follow an objective function that balances latency with incentive payouts  at regular or variable incentive budgets \cite{expect}. This mimics the free-rider problem by preventing high quality contributors from being represented \cite{incentive}.

\begin{table*}[ht]
\caption{Proposed benchmark health dataset and characteristic datatype, format, classes, and size. A base transfer learning model is given for each dataset and classification problem to be benchmarked.}
\centering
\def\arraystretch{1.1}
\begin{tabular}
{p{0.2\linewidth}p{0.20\linewidth}p{0.15\linewidth}p{0.1\linewidth}p{0.05\linewidth}p{0.10\linewidth}}
\hline
\textbf{Dataset} & \textbf{Data Type} & \textbf{Format} & \textbf{No. Classes} & \textbf{Size} & \textbf{Base Model}\\
\hline
NIH Chest X-Ray \cite{wang2017hospital} & Adult X-Ray Images & PNG & 13 & 25k & ResNet-34\\ 
    Fast MRI \cite{fast} & Brain Tissue Scans & 3D Arrays & 2 & 8k & U-Net\\
    PTB Dataset \cite{wagner2020ptb} & ECG Waveforms & Signal & 10 & $<$1k & ResNet-18\\ 
    MIMIC II \cite{mimic} & Clinical and Vital Signs & Tabular, Signal & 59 & 30k & XGBoost\\
\hline
\end{tabular}
\label{table:kysymys}
\end{table*}

\section{Experimental Design}

\subsection{Heterogeneous Client Distribution} 
Typical FL schemes assume client data is independent and identically distributed (IID) while real world environments typically possess high levels of heterogeneity \cite{iid}. We simulate non-IID clients by creating 10 data partitions and implementing a parameter $\alpha$ to simulate the percentage of client data that is non-IID for low, medium, and high values as depicted in Figure \ref{fig:fig2}. We also test on different maximum unique classes per client as either a single or biclass problem, depicted by notation Non-IID(1) and Non-IID(2). 

\subsection{AES Encryption} 
Advanced Encryption Standard-Galois Counter Mode (AES-GCM), a classic Authenticated Encryption mode with highly secure symmetric encryption, is incorporated in every federated round we train to protect privacy and client-server communication \cite{mcgrew2004security}. Parallel computing is a key component in our framework and the widely utilized AES-GCM is suitable due to the size and sensitive nature of healthcare-focused datasets \cite{dworkin2007recommendation}. Future work will extend this to a Trusted Execution Environment (TEE) to monitor model behavior in realistic privacy-preserving environments.

\section{Benchmark and Analysis} 
We demonstrate our methodology on the NIH Chest X-Ray Dataset by formulating a triclass problem of classifying  Pneumonia, non-Pneumonia condition, or no findings. Fairness approaches (see Table \ref{table:kysymys}) we account for are (1) the \textit{long-term client selection fairness constraint} and (2) two model schemes utilizing the \textit{reputation threshold contribution fairness}, one featuring an adversarial agent and one scheme without any agents. Classification models utilized a variable learning rate $\eta$ and fixed batch size $B$ per epoch $E$ for total clients $K$. For the NIH Chest X-Ray benchmark, we utilize $K$=10, $E$=1, $B$=64, and $\eta$ = 0.01 with decay rate = 0.992.

Our results demonstrate clear decreases in validation accuracy with increasing client heterogeneity, with different fairness mechanisms responding to variability differently. The reputation scheme with an additional adversarial agent outperforms the scheme without any agents at all levels of heterogeneity. Notably, the long term fairness constraint mechanism does not perform much differently than the standard federated scheme at low levels of heterogeneity but eventually converges to the optimal scheme as heterogeneity increases. 

\section{Conclusions and Future Work}
Federated learning in healthcare is a natural extension of standard FL architecture and is aligned with the healthcare domain focuses of security, privacy, and group fairness. In this paper we create a novel benchmark for FAFL mechanisms in healthcare contexts and propose augmenting it. Privacy preserving FAFL also brings challenges of communication overhead due to additional feedback loops and enforced security. These challenges will continue to be relevant as FAFL in healthcare evolves to more complex implementations. 

We propose future work to extend our analysis to the datasets listed in Table \ref{table:kysymys} and a wider range of mechanisms and schemes, including ones built to address heterogeneity such as FedProx. To further augment our analysis, we propose benchmarking estimators such as communication and execution times to further simulate real world settings.

\bibliographystyle{IEEEtran}
\bibliography{references}

\end{document}